**Controlled oxygen plasma treatment of single-walled carbon nanotube films improves osteoblast cells attachment and enhances their proliferation**

*Marie Kalbacova[a], Antonin Broz[a], Alexander Kromka[b], Oleg Babchenko[b], Martin Kalbac[c,d]\**

[a] Institute of Inherited Metabolic Disorders, First Faculty of Medicine, Charles University in Prague and General University Hospital in Prague, Ke Karlovu 2, 12852 Prague 2, Czech Republic

[b] Institute of Physics, ASCR, Cukrovarnická 10, 16253 Prague 6, Czech Republic

[c] J. Heyrovský Institute of Physical Chemistry, Academy of Sciences of the Czech Republic, Dolejškova 3, CZ-18223 Prague 8, Czech Republic

[d] Palacky Univ, Dept Phys Chem, Olomouc 77146, Czech Republic

**Abstract**

The effects of oxidative treatment of single-walled carbon nanotubes (SWCNTs) on the adhesion and proliferation of human osteoblasts (SAOS-2) were investigated. The surface properties of SWCNTs after oxygen plasma treatment were characterised by contact angle measurement, scanning electron microscopy and Raman spectroscopy. The immunofluorescent staining of vinculin, actin filaments and nuclei was used to probe cell adhesion and growth on SWCNT films. Our results show that adhesion and proliferation of human osteoblasts cultivated on SWCNT films indeed depends on the degree of an oxidative treatment. As an optimal procedure was found the treatment with oxygen plasma for 5 min. In the latter case the osteoblasts form a confluent layer with pronounced focal adhesions throughout the entire cell body. The optimal conditions compromise the effect of hydrophilic character of SWCNT films and the level of damage of SWCNT surface.

\*Corresponding author. Fax: +420 28658 2307. E-mail address: martin.kalbac@jh-inst.cas.cz (M.Kalbac)

# 1. Introduction

Due to their unique properties, single-walled carbon nanotubes (SWCNT) have many prospective applications, including biomedical engineering, drug delivery and medical chemistry. The significance of the toxicological issues associated with these materials has been the motivation for recent research. These studies have mostly focused on inhalation toxicology [1,2,3] or the effects on skin exposure [4,5]. There have also been studies on implanted carbon nanotubes [6] as well as *in vitro* studies investigating the influence of SWCNT on different cell cultures (endothelial cells [7], macrophages [8] or osteoblasts [9]).

It has been shown previously that functionalisation of carbon nanotubes dramatically changes their toxicity [10]. Functionalisation is usually performed by acid treatment of carbon nanotubes [11]. However, this method has several drawbacks since it is difficult to control this process. Strong acids are a potential risk for the environment if used in industry and also have the potential to contaminate the sample. On the other hand, treatment of SWCNTs with a dry oxygen plasma seems to be an economically reasonable and effective technique for the modification of the surface of carbon nanotubes [12,13] due to short reaction times and simplicity of the process. This process is also clean, and therefore it can be applied to products at the end of manufacturing process. Nevertheless, the influence of oxygen plasma treatment of SWCNT on biological systems is still unknown.

Osteoblasts are anchorage-dependent cells which form bone. Therefore, adhesion of osteoblasts to a substrate is a crucial prerequisite for subsequent cell functions such as proliferation, protein synthesis (e.g. extracellular matrix proteins, morphogenic factors and osteoinductive molecules) and formation of mineral deposits. The cell adhesion is primarily mediated by integrins, a widely expressed family of transmembrane adhesion receptors [14]. Upon ligand binding, integrins rapidly associate with the actin cytoskeleton due to binding to adaptor proteins and cluster together to form focal adhesions, which are discrete complexes that contain structural and signaling molecules [15]. Focal adhesions are central elements in the adhesion process, functioning as structural links between the cytoskeleton and the extracellular matrix to mediate stable adhesion and migration. It could be distinguished between peripheral adhesion sites (focal complexes) enriched in integrin αVβ3 which facilitate protrusion formation and more centrally located integrin α5β1- enriched fibrillar adhesion sites essential for extracellular matrix (ECM) assemble and stable cell attachment [16]. Furthermore, in combination with growth factor receptors, focal adhesions activate signaling pathways that regulate transcription factor activity and direct cell growth and differentiation [17]. Vinculin is one of the most important adaptor proteins specifically associated with focal adhesions responsible for actin attachment to the plasma membrane and it is present in a wide variety of cell types [18]. Therefore, immunelabeling of vinculin by means of a specific antibody provides a crucial and defined detection system for the localization and size of focal adhesions, which give an information of the influence of surface properties on cell growth (substrate compliance – rigid or soft, ligand spacing related to specific ECM protein presence on the surface, surface topography, surface free energy, etc.)[19].

Here, we used immunofluorescent staining of vinculin, actin filaments and nuclei to probe cell adhesion and growth on SWCNT films treated by dry oxygen plasma. By altering the duration of exposure of SWCNT films to oxygen plasma, we prepared samples with different levels of damage and wettability. We found optimal conditions which compromise competing effects of hydrophilic character and damage of the SWCNT substrate.

## 2. Experimental

*SWCNTs*

SWCNT films were prepared as follows: HiPco Super purified nanotubes were sonicated in N-methylpyrrolidone, then filtered through a membrane filter (0.2 µm) and washed with an excess of ethanol and water. The film on the membrane filter was stamped onto a glass substrate. Finally, the samples were heated at 450°C in a flow of argon and hydrogen for 30 min. Before cell experiments, SWCNT films were sterilised for 10 min in 70% ethanol, washed in PBS and allowed to dry.

*Oxygen plasma treatment*

The oxygen plasma treatment was performed by radio-frequency maintained plasma (13.56 MHz) in a double plasma source system (AK 400, Roth & Rau). The process was performed at room temperature with the following parameters: pressure 0.35 mbar, rf power 50 W and self bias voltage 8 V, oxygen flow 50 sccm. The rf plasma treatment times were 1, 5 and 30 minutes.

*Contact angle (CA) measurement*

The surface wettability of SWCNTs was calculated from water droplet contact angle measurements. The CA measurements were obtained at room temperature by a static method in a material–water droplet system using a reflection goniometer (Surface Energy Evaluation (SEE) System). 3 µl of deionised water were dispersed on the SWCNT sample surfaces and the formed drop was captured by a digital CCD camera. The contact angle was calculated by a multipoint fitting of the drop profile using SEE software.

*Raman spectroscopy*

The Raman spectra were excited by mixed $Ar^+/ Kr^+$ laser (Innova 70C series, Coherent) and were recorded by a Labram HR spectrometer (Horiba Jobin Yvon) interfaced to an Olympus BX-41 microscope (objective 50x). The laser power impinging on the cell window or on the dry sample was between 0.1 and 1 mW. The spectrometer was calibrated using the $F_{1g}$ mode of Si at 520.2 cm$^{-1}$.

*SEM imaging*

SEM imaging was performed using a high resolution SEM S-4800 (Hitachi).

*Cells*

SAOS-2, a human osteoblast-like cell line, was obtained from DSMZ (Deutsche Sammlung von Mikroorganismen und Zellkulturen GmbH). Cells were grown at 37°C in 5% $CO_2$ in McCoy's 5A medium without phenol red (BioConcept) supplemented with 15% heat inactivated fetal bovine serum (Biowest), penicillin (20 U/ml) and streptomycin (20 µg/ml). Cells were plated (25 000 cells/cm$^2$) on glass substrates covered with pristine SWCNT films (P) and films after oxygen-plasma treatment for 1 minute (O1), 5 minutes (O5) and 30 minutes (O30) and incubated for 48 h under the tissue culture conditions.

*Fluorescent staining of cells*

The focal adhesions of SAOS-2 cells were characterised morphologically by immunofluorescent staining of vinculin (1:150, Sigma, anti-mouse Alexa 568). Visualisation of the cytoskeleton was performed by staining of actin filaments (phalloidin-Alexa-488, 1:100, Invitrogen) and nuclei were visualised with 4',6-diamidino-2-phenylindole (DAPI, 1:1000, Sigma). An epi-fluorescent Nikon E-400 microscope was used (Hg lamp, Uv-2A, B-2A and G-2A filter set) and data were recorded by DS-5M-U1 Colour Digital Camera (Nikon).

*Cell number*

To count cells, twelve size-calibrated fluorescent pictures of DAPI-stained cell nuclei from each surface type were obtained using a Nikon E400 microscope with a 4x lens. Areas of 1 mm$^2$ were cut out of these calibrated pictures and the cell number was counted using NIS Elements software (LIM).

*Focal adhesion size analysis*

Fluorescent images of cells taken by the epi-fluorescent Nikon E-400 microscope were analysed for the size of focal adhesions using NIS Elements (LIM). The long and short axes of focal adhesions were determined. For statistical analysis, ANOVA was used.

## 3. Results and discussion

SWCNTs were deposited on glass substrates to form a compact film and subsequently exposed to oxygen plasma for different times. Figure 1 shows SEM pictures of a pristine sample (P; Fig. 1A) and samples after different oxygen plasma treatment for 1 min, 5 min and 30 min (O1, O5, O30; Fig.1 B-D, respectively). Obviously, prolonged oxygen plasma treatment resulted in damage to the SWCNT film surface, which became rougher as the treatment time increased. The oxygen plasma also caused a decrease in the conductivity of the sample. The practical consequence of this was a stronger charging of the oxygen plasma-treated samples during SEM imaging. The decrease of conductivity was in agreement with previous results [20] and it may have been caused both by a decrease of the intrinsic conductivity of the nanotubes and by higher contact resistance between the nanotubes. The longest duration of exposure of SWCNT films to oxygen plasma (30 min) resulted in significant morphological changes. On the other hand, oxygen plasma exposure times of 1 minute and 5 minutes resulted in no obvious removal and/or extensive degradation of the SWCNT.

Changes in the surface properties of oxygen plasma-treated samples were further assessed by contact angle measurements. The compact layer of pristine SWCNTs exhibited a hydrophobic surface with a contact angle α = 100°±5°. As the SWCNTs were treated by oxygen plasma for 1 min (O1), the surface wettability changed drastically from hydrophobic to hydrophilic with a characteristic contact angle α=21°±5°. Prolonging the oxygen plasma treatment to 5 min (O5) resulted in a further decrease of the contact angle down to 11°±3°. Extensive oxygen plasma treatment, i.e. for 30 min (O30), did not further decrease the contact angle which was already saturated at a value of 12°±3°.

We noted that the process time required for achieving hydrophilic surfaces could be reduced to 10 seconds if higher pressure and higher rf power was used (results not shown here). However, such process conditions are more difficult to control.

Previous XPS measurements performed on plasma-treated multi-wall carbon nanotubes (MWCNTs) have shown that the amount of oxygen-containing functional groups increased with treatment time [12]. Hence, the abrupt change in wettability from a hydrophobic to a hydrophilic character was obviously caused by an increased amount of oxygen-containing functional groups. We assume that plasma treatment first initializes defect formation and openings in the SWCNT wall and these new surfaces behave as reactive sites for oxygen bonding.

Figure 2 shows the Raman spectra of the pristine and the oxygen plasma treated samples. All Raman spectra exhibited typical features of SWCNTs: the radial breathing mode (RBM), the tangential mode (G), the defect induced mode (D) and the double resonant feature (G' mode).

The Raman spectra changed significantly after oxygen plasma treatment. The oxygen plasma treatment changed the overall intensity of the Raman spectra (note that the Raman spectrum of the pristine sample is scaled by a factor of 0.5 in Figure 2). There were also subtler changes in the frequency of some Raman features and an obvious increase in the intensity of the D mode.

The RBM region of the pristine sample contained several Raman bands between 150 and 350 cm$^{-1}$ that reflected a relatively broad diameter distribution of the HiPco tubes used in our experiment. After oxygen plasma treatment, the intensity of the RBM bands was reduced; hence, only the most intensive bands at 220 and 260 cm$^{-1}$ were clearly distinguishable in the Raman spectra. Furthermore, the frequency of the RBM bands varied slightly in the oxygen plasma-treated samples. We attributed these changes in frequency to the change of the resonant condition, as has been suggested recently [21] due to oxygen-induced doping. We also could not exclude a preferential etching of tubes with specific diameters by oxygen plasma.

The G mode (found between 1500 and 1600 cm$^{-1}$) consisted of two bands. The lower frequency component of the G mode was narrow; hence, mostly semiconducting tubes contributed to the Raman spectra shown in Fig. 2. The shape and frequency of the G mode were not changed significantly after oxygen plasma treatment which means that the doping of the sample was only weak, if it occurred at all [22,23].

In contrast to the other Raman features, the intensity of the D mode was significantly increased after oxygen plasma treatment of the samples. The increased D mode intensity suggested the formation of new defects. Nevertheless, it has been shown recently that the D mode intensity also depends on the doping level [24]. Doping could not be excluded in our sample, hence the D mode intensity should be related to the other Raman features (the G or the G' band). The changes to $I_D/I_G$ and $I_D/I_{G'}$ are frequently used for the evaluation of defects in carbon nanotube samples. It should be noted that it is not preferable to use peak height for the calculation of $I_D/I_G$ and $I_D/I_{G'}$ ratios. This is because peak width can be also significantly influenced by doping [24]. Hence, both the $I_D/I_G$ and $I_D/I_{G'}$ could be different for different doping levels in the sample, even if the amount of defects was not changed by doping [24]. It is more reliable to use peak area for the calculation of $I_D/I_G$ and $I_D/I_{G'}$ ratios. The area of the peaks is also changed by doping; however, it has been demonstrated that the $I_D/I_G$ and $I_D/I_{G'}$ ratios do not change with doping if the integrated area of the peaks is used. The calculated $I_D/I_G$ ratios (using integrated areas) for the studied samples were 0.10 (P), 0.58 (O1), 0.78 (O5) and 0.45 (O30). The $I_D/I_{G'}$ ratios were found to be 0.21 (P), 1.52 (O1), 2.37 (O5) and 1.00 (O30). According to our expectations, both the $I_D/I_G$ and $I_D/I_{G'}$ ratios increased with oxygen plasma treatment compared to the pristine sample. The ratios were also higher for the 5 minutes treated sample than for the 1 minute treated sample. This confirmed that oxygen plasma creates defects, such that a longer the duration of oxygen plasma treatment is associated with the creation of more defects. Nevertheless, for the longest treatment time of 30 minutes (O30), we observed decreased $I_D/I_G$ and $I_D/I_{G'}$ ratios. A decrease in $I_D/I_G$ and $I_D/I_{G'}$ ratios with the creation of defects has also been observed by others [25]. The reason for this is that each defect influences the Raman spectra of a certain area of nanotube. If the defect density exceeds a certain level, it is probable that in the area influenced by one defect is present another one. In such cases, the contribution to the D band is weaker [25].

The pristine and oxygen plasma-treated samples were used as substrates for the growth of human osteoblasts. The osteoblasts cultivated on SWCNT substrates showed a distinct pattern after 48 h (Fig. 3). Cells cultivated on the pristine sample and on the sample treated with oxygen plasma for 1 minute (Fig. 3A and 3B, respectively) developed a layer of cells of about 65% confluence. Cells cultivated on the SWCNT layer treated with oxygen plasma for 5 minutes were almost confluent after 48 h of incubation and most of the cells were in contact, forming a compact layer (Fig. 3C). The 30 min oxygen plasma treatment of SWCNT caused cells to grow mostly solitarily, avoiding contact with each other. To confirm our conclusions from the visual analysis of the fluorescent picture, we also counted the number of cells on each sample shown on Fig. 4. It was obvious that on a counted area of 1 mm$^2$, the cell number on the O5 sample was almost double compared to the other samples (P, O1, and O30). This demonstrated normal cell growth (with doubling time of 44 h for this cell line) on O5 substrate and almost no growth comparable on the other surfaces (P, O1, and O30). Furthermore, a confluent layer was formed on the O5 sample while a continuous layer of cells was found on the P and O1 samples and solitarily growing cells were found on the O30 sample.

Surface wettability may have affected the proliferation of cells since the initial phase of attachment involves physical/chemical linkages between cells and surfaces through ionic forces, or indirectly through an alteration in the adsorption of conditioning molecules, e.g. proteins. In this study, the surface wettability was significantly different for the pristine and oxygen-plasma treated samples. As shown for other carbon substrates such as nanocrystalline diamond [26], when the only substrate specific factor is the wettability, the hydrophilic substrate is more suitable for the cell growth. (Osteoblasts can still grow on hydrophobic substrates, but the rate of growth is strongly reduced [26].) Hence, the contact angle of 21° for the O1 sample and about 12° for the O5 and O30 samples (hydrophilic character) should be more convenient for cell growth than the hydrophobic surface of the P sample (contact angle about 100°). In our case, surprisingly, the number of cells on the hydrophilic O1 sample was significantly lower ($p<0.05$) than on the hydrophobic P sample and the cell number on the O30 sample was comparable to that on the P and O1 samples. A significantly higher cell number was found only on the O5 sample (almost double in comparison to other samples). Consequently, this experiment clearly shows that the wettability of SWCNTs films is not the only factor responsible for osteoblasts adhesion and growth, but the surface topography (level of damage in the SWCNT layer) or other factors (surface free energy, surface charge, etc.) also seemed to be important.

The unique surface properties of nano-phase materials, namely a higher number of atoms at the surface compared to bulk, greater areas of surface defects (such as edge/corner sites and particle boundaries) and a larger proportion of surface electron delocalisation may influence initial protein interactions (such as with proteins originating from the fetal bovine serum included in the culture medium) that further control cell adhesion [27,28], a determinant event for subsequent cell proliferation and function. Carbon nanotubes might adsorb a large number of proteins due to their larger surface area and unique electronic, catalytic and chemical properties [29]. The different modifications to the SWCNTs by oxygen plasma may have had a strong effect on protein adsorption and thus affect the cell adhesion and proliferation. It has already been reported [30] that competitive protein adsorption at a bioactive surface varies in three ways, which are the quantity of protein adsorbed, the species of protein adsorbed and the conformation of the adsorbed protein. It should be noted that carbon nanotubes can adsorb also important nutrients (like folic acid) from cell culture medium [31], which may also influence cell behavior. However, we believe that the results of our study are not significantly influenced by the latter effect since we used relatively short incubation time (48 h). In addition we do not use biochemical methods (determination of dehydrogenases function in mitochondria using MTS test). In the case of determination of cellular adhesion, protein influence seems to be more relevant. The combination of the surface properties of the O5 substrate (hydrophilic character, rougher surface, specific amount of oxygen atoms) may have been optimal for protein adsorption in a functional conformation or for optimal protein selection. The importance of quantitative protein adsorption was demonstrated previously on a $TiO_2$ substrate [32] where the better adhesion of the cells was observed for hydrophilic substrate due to a higher amount of cell-adhesion-mediating proteins compared to hydrophobic substrate.

For a more detailed analysis of cell adhesion to the different substrates, we also evaluated fluorescent images of focal adhesions. Fluorescent staining of vinculin, a structural protein involved in the formation of focal adhesions, shows that osteoblasts cultivated on samples treated differently with oxygen plasma used a distinct pattern of adhesion. Osteoblasts cultivated on pristine SWCNTs (Fig. 4A) and on the O1 substrate (Fig. 4B) formed focal adhesions on the periphery of the cell with only a few adhesions under the cell center. On the other hand, cells cultivated on the O5 substrate adhered with the entire cell body and with relatively large adhesion plaques and sometimes even with adhesion fibers (Fig. 4C). A totally different adhesion pattern was obtained in cells cultivated on the O30 substrate. Cells formed only small adhesions, the periphery of the cells was very irregular and most of the structural protein vinculin was in a diffused form within the cytoplasm.

To strengthen the data from the fluorescent images, we used an image analysis program (NIS Elements, LIM) to obtain the size of adhesions. We graphically compared the size of the long and short axis for each focal adhesion in cells cultivated on the different substrates (Fig. 6 and 7). From the graphs, it was clear that adhesions formed on the O30 substrate were very small on both axes compared to the adhesions formed on the other substrates. Focal adhesions formed on the O5 substrate were of the same length but narrower than the adhesions formed on the P substrate. On the other hand, focal adhesions formed on the O1 sample were comparable to the P substrate. Despite the fact that we analysed a large number of cells on each substrate, the variability in focal adhesion size parameters was very high, thus the real differences could have been lost. Nevertheless, the multiple comparison procedures confirmed the significant difference of the long axis (at the 0.01 level) between the O30 substrate and the all other substrates. For the short axis the tests confirmed the significant difference between the O30 substrate and all other tested substrates and also between O5 substrate and P substrate.

Vinculin potentially serves as a stabilising protein in focal adhesions, therefore the amount of vinculin present may be indicative of the motility of a cell on a given substrate [33]. Focal adhesions play opposing roles in cell motility, aiding both in the generation of cellular strain to generate lamellipodium formation and polarised motility, and as anchoring complexes that resist detachment from the substrate and aid in cellular spreading. Focal adhesions are most developed and numerous in flattened cells [34]. It has also been noted that the adhesion strength of flattened cells to the substrate was also greater compared to unflattened cells [35]. Thus, from our results it could be suggested that cells cultivated on the O5 substrate which formed large and numerous focal adhesions under the entire cell body were strongly adhered to the substrate, which is a prerequisite for increased proliferation and possibly for differentiation. On the other hand, osteoblasts on the pristine and O1 substrates formed focal adhesions only on the periphery of the cells and revealed weaker adhesion, causing slower cell proliferation with possibly increased cell motility. The tiny focal adhesions and mostly diffused vinculin in cells incubated on the O30 substrate indicated that prolonged plasma treatment of SWCNT films results in a poor osteoblast cells adhesion and thus also limits proliferation and differentiation of the cells.

An increased amount of oxygen and topographical changes to the SWCNT film influenced the attachment, growth and differentiation of living cells. Subsequently, the cell number and adhesion patterns varied in regard to the treatment time which influenced surface properties, i.e. the degree of wettability, morphology, topography and mechanical stability. Generally, oxygen-rich surfaces are often hydrophilic and oxidised diamond surfaces ( as an example of another carbon material) were found to be preferable surfaces for cell growth [36,37]. The observed effect was assigned to the absorption selectivity of proteins from the fetal bovine serum used in the growth medium (fibronectin, vitronectin, bovine serum albumin, etc.). It was also found that proteins adsorbed in about the same monolayer thickness (2–4 nm) on both diamond surfaces (C-H and C-O), but in different conformations. However, in our case, applying a hydrophilic surface did not follow the trend described above. But another factor which plays an important role in cell cultivation is the geometry and topography of the substrate. In our other studies, we have found that hierarchically structured diamond surfaces have a significant influence on the adhesion and growth of SAOS-2 cells [38,39]. The comparison of SWCNT and diamond surfaces is interesting in this case since: a) the same type of cell line was used (SAOS-2), b) both surfaces were carbon-based nano-phase materials and c) the surface wettability was very similar in both materials (i.e. hydrophilic surfaces with wetting angle around 20°).

In the first assumption, the oxygen plasma treated SWCNTs in this paper represent a similar situation as the structured diamond surface. Prolonged oxygen plasma treatment changed the surface morphology (as showed by the SEM images in Fig. 1). The additional influence on cell growth was the total contact area of the cultivated cells with the substrate. Once the top diamond morphology was patterned to nano-rods, only tiny adhesions of SAOS-2 cells were observed [40]. We propose that 30 min plasma treatment of the SWCNT layer resulted in the formation of a tiny, randomly-distributed nano-structured morphology. However, the O5 substrate formed a compact layer of nanostructured SWCNT film, reminiscent of a bone-like structure which is familiar to osteoblasts (bone forming cells). It is known that for effective integrin clustering and proper formation of focal adhesion is critical ligand spacing under 70 nm (RGD ligand for integrin receptor) and that the cell adhesion is turned-on on disordered ligand patterns and turned-off on ordered patterns [41]. Thus we suggest that surface obtained after 30 min plasma treatment (partially destroyed) does not provide a topography required for proper protein binding and conformation, thus the cells cannot form stable focal adhesions. On the other hand, surface treated only for 5 min by oxygen plasma changes its surface topography enabling proper protein adsorption in a active conformation and thus allowing formation of stable focal adhesions. Another factor influencing cell growth could be the mechanical stability of the substrate. The oxygen plasma treatment leads presumably to the formation of the functional groups. It was also shown previously that functionalized carbon nanotubes change both wettability and also *debundeling* tendency [42]. In addition as can be seen from Raman spectra (Fig. 2) the O-plasma treatment creates defects and it also leads to the morphological changes of the nanotubes film surface (SEM images, Figure 3). Based on these results we suppose that nanotubes can be shortened and became less compact. Thus, oxygen plasma treated SWCNT may imitate a soft substrate material, which result in a less stable cell adhesion to the substrate as observed for O30 sample.

## 4. Conclusion

We tested the influence of oxygen plasma treatment applied to SWCNT films on osteoblasts adhesion and their growth on these films. A dry treatment of SWCNT film in oxygen plasma at low rf power resulted in the formation of defects over the SWCNT film, as confirmed by Raman measurements. Contact angle measurements showed that these newly-initialised defects changed the hydrophobic character of the pristine SWCNT surface into a hydrophilic one. Furthermore, we investigated the topography of oxygen plasma-treated samples by SEM. Significant changes in the topography of the SWCNT surfaces were found only in the extensively-treated SWCNT sample (30 min). The SWCNT sample treated for 5 min was found to be the best surface for osteoblast adhesion accompanied by cell proliferation. On the other hand, strong changes in the topography of the sample treated with oxygen plasma for 30 min reduced cell adhesion, despite its hydrophilic character. Our results demonstrate that both hydrophilic character and specific surface topography are important factors for cell adhesion and proliferation. Therefore, the control of these factors is crucial for the future design of implants and the application of new materials in regenerative medicine.

**Acknowledgement**

This work was supported by the Academy of Sciences of the Czech Republic (contracts Nos., IAA400400911, KAN400100701, KAN115600801 and KAN200100801), Czech Grant agency (P204/10/1677) and by the Czech Ministry of Education, Youth and Sports (contracts No. ME9060 and MSM 0021620806). M. Kalbacova acknowledges the support from 2010 L'Oreal-UNESCO For Women in Science Fellowship. The authors are gratefull to B. Sediva (University of West Bohemia, Pilsen, Czech Republic) for the help with statistical analysis of the data.

Reference List

1. Muller J, Huaux F, Lison D. Respiratory toxicity of carbon nanotubes: How worried should we be? Carbon 2006;44(6):1048-1056.

2. Grubek-Jaworska H, Nejman P, Czuminska K, Przybylowski T, Huczko A, Lange H, Bystrzejewski M, Baranowski P, Chazan R. Preliminary results on the pathogenic effects of intratracheal exposure to one-dimensional nanocarbons. Carbon 2006;44(6):1057-1063.

3. Warheit DB, Laurence BR, Reed KL, Roach DH, Reynolds GA, Webb TR. Comparative pulmonary toxicity assessment of single-wall carbon nanotubes in rats. Toxicol.Sci. 2004;77(1):117-125.

4. Monteiro-Riviere NA, Nemanich RJ, Inman AO, Wang YY, Riviere JE. Multi-walled carbon nanotube interactions with human epidermal keratinocytes. Toxicol.Lett. 2005;155(3):377-384.

5. Shvedova AA, Castranova V, Kisin ER, Schwegler-Berry D, Murray AR, Gandelsman VZ, Maynard A, Baron P. Exposure to carbon nanotube material: Assessment of nanotube cytotoxicity using human keratinocyte cells. Journal of Toxicology and Environmental Health-Part A 2003;66(20):1909-1926.

6. Koyama S, Endo M, Kim YA, Hayashi T, Yanagisawa T, Osaka K, Koyama H, Haniu H, Kuroiwa N. Role of systemic T-cells and histopathological aspects after subcutaneous implantation of various carbon nanotubes in mice. Carbon 2006;44(6):1079-1092.

7. Flahaut E, Durrieu MC, Remy-Zolghadri M, Bareille R, Baquey Ch. Investigation of the cytotoxicity of CCVD carbon nanotubes towards human umbilical vein endothelial cells. Carbon 2006;441093-1099.

8. Fiorito S, Serafino A, Andreola F, Bernier P. Effects of fullerenes and single-wall carbon nanotubes on murine and human macrophages. Carbon 2006;441100-1105.

9. Chlopek J, Czajkowska B, Szaraniec B, Frackowiak E, Szostak K, Beguin F. In vitro studies of carbon nanotubes biocompatibility. Carbon 2006;441106-1111.

10. Liu Z, Tabakman S, Welsher K, Dai HJ. Carbon Nanotubes in Biology and Medicine: In vitro and in vivo Detection, Imaging and Drug Delivery. Nano Research 2009;2(2):85-120.

11. Zhang XF, Sreekumar TV, Liu T, Kumar S. Properties and structure of nitric acid oxidized single wall carbon nanotube films. J.Phys.Chem.B 2004;108(42):16435-16440.

12. Xu T, Yang JH, Liu JW, Fu Q. Surface modification of multi-walled carbon nanotubes by O-2 plasma. Applied Surface Science 2007;253(22):8945-8951.

13. Chirila V, Marginean G, Brandl W. Effect of the oxygen plasma treatment parameters on the carbon nanotubes surface properties. Surface & Coatings Technology 2005;200(1-4):548-551.

14. Hynes RO. Integrins: bidirectional, allosteric signaling machines. Cell 2002;110(6):673-687.

15. Geiger B, Bershadsky A, Pankov R, Yamada KM. Transmembrane crosstalk between the extracellular matrix--cytoskeleton crosstalk. Nat.Rev.Mol.Cell Biol. 2001;2(11):793-805.

16. Papusheva E, Heisenberg CP. Spatial organization of adhesion: force-dependent regulation and function in tissue morphogenesis. Embo Journal 2010;29(16):2753-2768.

17. Giancotti FG, Ruoslahti E. Integrin signaling. Science 1999;285(5430):1028-1032.

18. Geiger B. 130K Protein from Chicken Gizzard - Its Localization at the Termini of Microfilament Bundles in Cultured Chicken-Cells. Cell 1979;18(1):193-205.

19. Geiger B, Spatz JP, Bershadsky AD. Environmental sensing through focal adhesions. Nat.Rev.Mol.Cell.Bio. 2009;10(1):21-33.

20. Kim S, Kim HJ, Lee HR, Song JH, Yi SN, Ha DH. Oxygen plasma effects on the electrical conductance of single-walled carbon nanotube bundles. Journal of Physics D-Applied Physics 2010;43(30):

21. Kalbac M, Farhat H, Kavan L, Kong J, Sasaki K, Saito R, Dresselhaus MS. Electrochemical charging of individual single-walled carbon nanotubes. Acs Nano 2009;3(8):2320-2328.

22. Kalbac M, Farhat H, Kavan L, Kong J, Dresselhaus MS. Competition between the Spring Force Constant and the Phonon Energy Renormalization in Electrochemically Doped Semiconducting Single Walled Carbon Nanotubes. Nano Letters 2008;8(10):**3532**-3537.

23. Kalbac M, Kavan L, Dunsch L, Dresselhaus MS. Development of the tangential mode in the Raman spectra of SWCNT bundles during electrochemical charging. Nano Letters 2008;8(4):1257-1264.

24. Kalbac M, Kavan L. The influence of doping on the Raman intensity of the D band in single walled carbon nanotubes. Carbon 2010;**48**832-838.

25. Lucchese MM, Stavale F, Ferreira EHM, Vilani C, Moutinho MVO, Capaz RB, Achete CA, Jorio A. Quantifying ion-induced defects and Raman relaxation length in graphene. Carbon 2010;48(5):1592-1597.

26. Kalbacova M, Kalbac M, Dunsch L, Kromka A, Vanecek M, Rezek B, Hempel U, Kmoch S. The effect of SWCNT and nano-diamond films on human osteoblast cells. Phys.Status Solidy B 2007;244(11):4356-4359.

27. Webster TJ, Ergun C, Doremus RH, Siegel RW, Bizios R. Specific proteins mediate enhanced osteoblast adhesion on nanophase ceramics. Journal of Biomedical Materials Research 2000;51(3):475-483.

28. Webster TJ, Ergun C, Doremus RH, Siegel RW, Bizios R. Enhanced functions of osteoblasts on nanophase ceramics. Biomaterials 2000;21(17):1803-1810.

29. Li XM, Gao H, Uo M, Sato Y, Akasaka T, Abe S, Feng QL, Cui FZ, Watari F. Maturation of osteoblast-like SaoS2 induced by carbon nanotubes. Biomedical Materials 2009;4(1):015005

30. Hing KA. Bioceramic bone graft substitutes: Influence of porosity and chemistry. International Journal of Applied Ceramic Technology 2005;2(3):184-199.

31. Guo L, Bussche AV, Buechner M, Yan AH, Kane AB, Hurt RH. Adsorption of essential micronutrients by carbon nanotubes and the implications for nanotoxicity testing. Small 2008;4(6):721-727.

32. Sousa SR, Moradas-Ferreira P, Barbosa MA. TiO2 type influences fibronectin adsorption. Journal of Materials Science-Materials in Medicine 2005;16(12):1173-1178.

33. Alenghat FJ, Fabry B, Tsai KY, Goldmann WH, Ingber DE. Analysis of cell mechanics in single vinculin-deficient cells using a magnetic tweezer. Biochemical and Biophysical Research Communications 2000;277(1):93-99.

34. Hunter A, Archer CW, Walker PS, Blunn GW. Attachment and Proliferation of Osteoblasts and Fibroblasts on Biomaterials for Orthopedic Use. Biomaterials 1995;16(4):287-295.

35. Lotz MM, Burdsal CA, Erickson HP, Mcclay DR. Cell-Adhesion to Fibronectin and Tenascin - Quantitative Measurements of Initial Binding and Subsequent Strengthening Response. Journal of Cell Biology 1989;109(4):1795-1805.

36. Kalbacova M, Michalikova L, Baresova V, Kromka A, Rezek B, Kmoch S. Adhesion of osteoblasts on chemically patterned nanocrystalline diamonds. Phys.Status Solidy B 2008;245(10):2124-2127.

37. Rezek B, Ukraintsev E, Kromka A, Ledinsky M, Broz A, Noskova L, Hartmannova H, Kalbacova M. Assembly of osteoblastic cell micro-arrays on diamond guided by protein pre-adsorption. Diamond and Rel.Mat. 2010;19(2-3):153-157.

38. Kalbacova M, Rezek B, Baresova V, Wolf-Brandstetter C, Kromka A. Nanoscale topography of nanocrystalline diamonds promotes differentiation of osteoblasts. Acta Biomater. 2009;5(8):3076-3085.

39. Kromka A, Rezek B, Kalbacova M, Baresova V, Zemek J, Konak C, Vanecek M. Diamond Seeding and Growth of Hierarchically Structured Films for Tissue Engineering. Advanced Engineering Materials 2009;11(7):B71-B76.

40. Kalbacova M, Broz A, Babchenko O, Kromka A. Study on cellular adhesion of human osteoblasts on nano-structured diamond films. Phys.Status Solidy B 2009;246(11-12):2774-2777.

41. Huang JH, Grater SV, Corbellinl F, Rinck S, Bock E, Kemkemer R, Kessler H, Ding JD, Spatz JP. Impact of Order and Disorder in RGD Nanopatterns on Cell Adhesion. Nano Letters 2009;9(3):1111-1116.

42. Ruelle B, Peeterbroeck S, Gouttebaron R, Godfroid T, Monteverde F, Dauchot JP, Alexandre M, Hecq M, Dubois P. Functionalization of carbon nanotubes by atomic nitrogen formed in a microwave plasma Ar+N-2 and subsequent poly(epsilon-caprolactone) grafting. Journal of Materials Chemistry 2007;17(2):157-159.

**Figures and captions:**

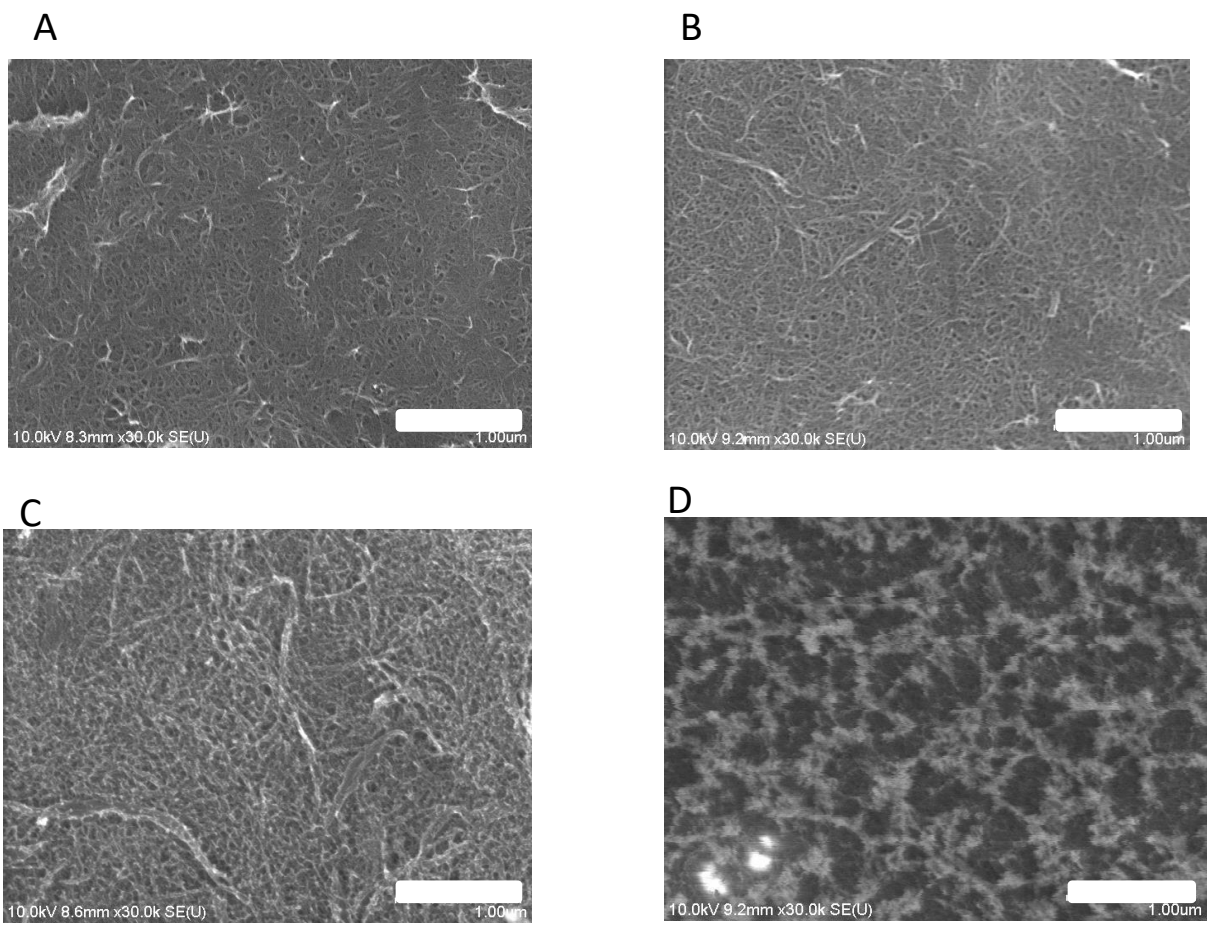

**Figure 1.** SEM images of pristine sample (A) and samples after oxygen-plasma treatment: 1 min (B), 5 min (C), 30 min (D). The scale bar is 1 um.

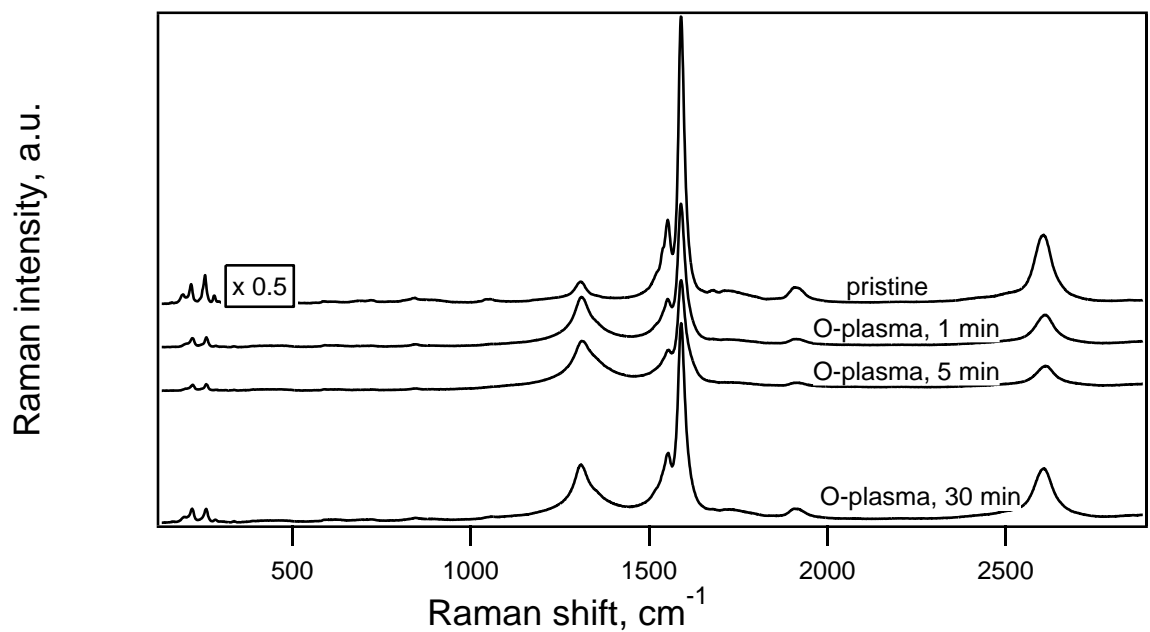

**Figure 2.** Raman spectra of pristine sample and samples after oxygen plasma treatment: 1 min, 5 min, 30 min (top to bottom). The Raman spectra were excited by 633 nm laser excitation energy. The scale is same for all spectra except pristine sample which is scaled by factor 0.5.

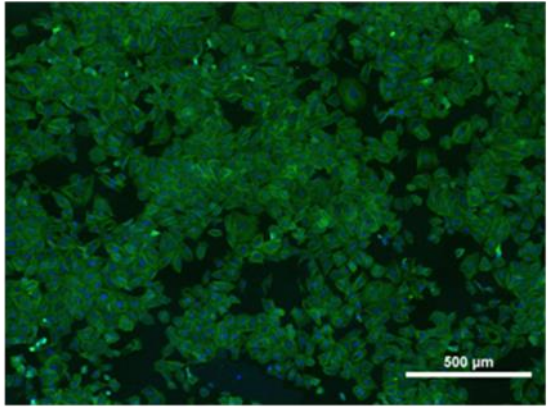
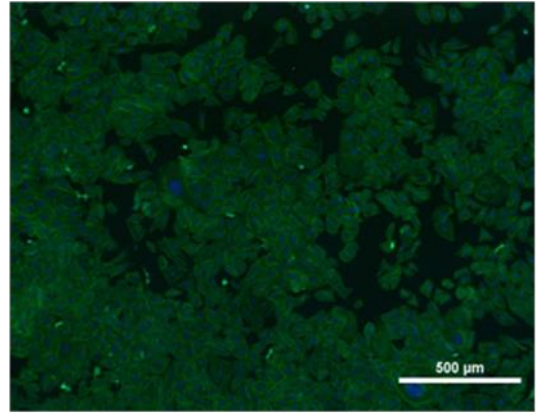
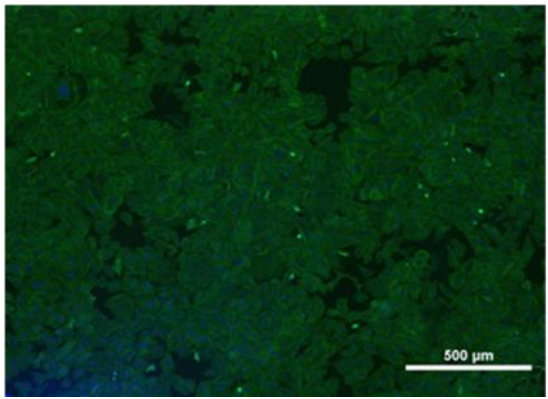
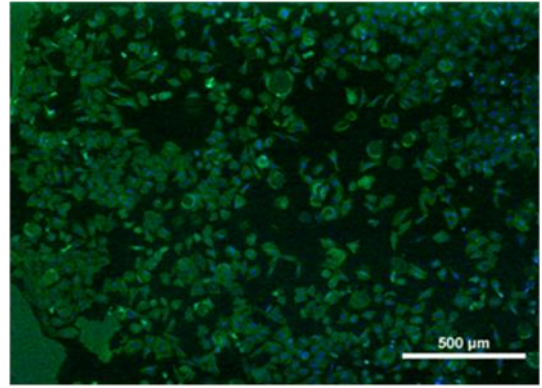

**Figure 3.** Fluorescent images of osteoblasts (SAOS-2) cultivated for 48 h on pristine sample (A) and samples after oxygen plasma treatment: 1 min (B), 5 min (C), 30 min (D). The scale bar is 500 um.

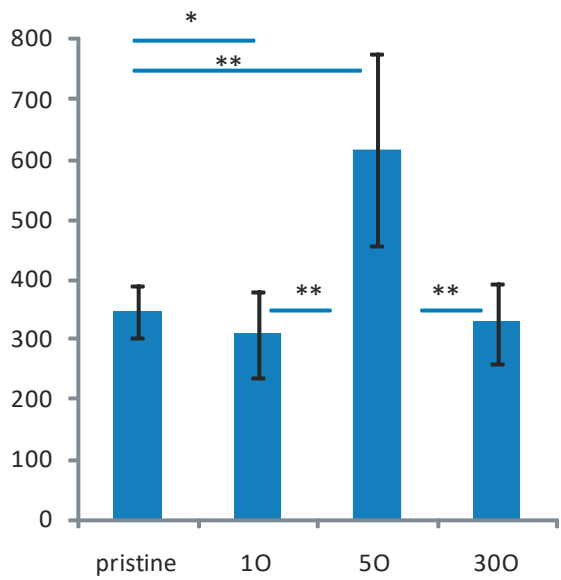

**Figure 4.** Cell number on different SWCNT types on area of 1mm$^2$,*p<0.05,**p<0.01

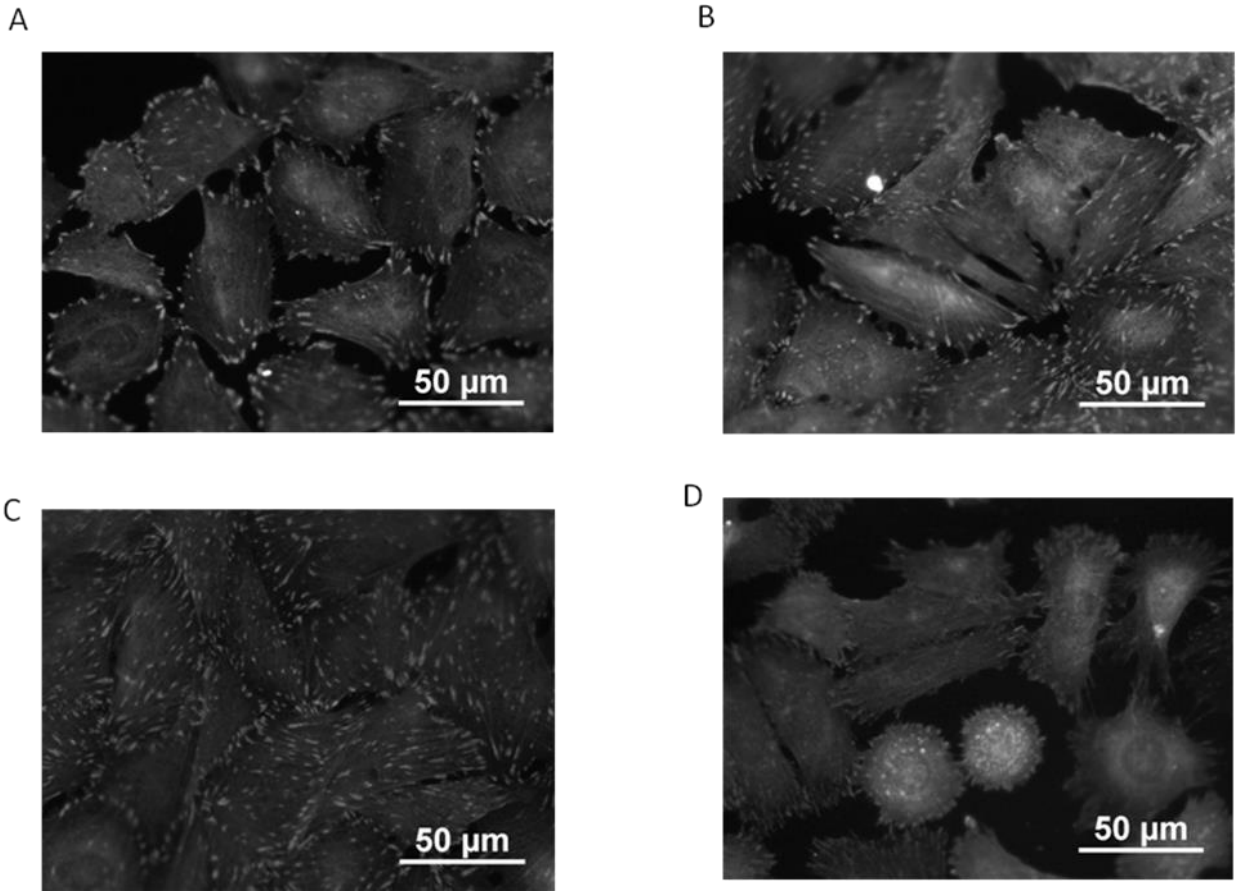

**Figure 5.** Fluorescent images of vinculin (structural focal adhesion protein) in osteoblasts cultivated for 48 h on pristine sample (A) and samples after oxygen plasma treatment: 1 min (B), 5 min (C), 30 min (D). The scale bar is 50 um.

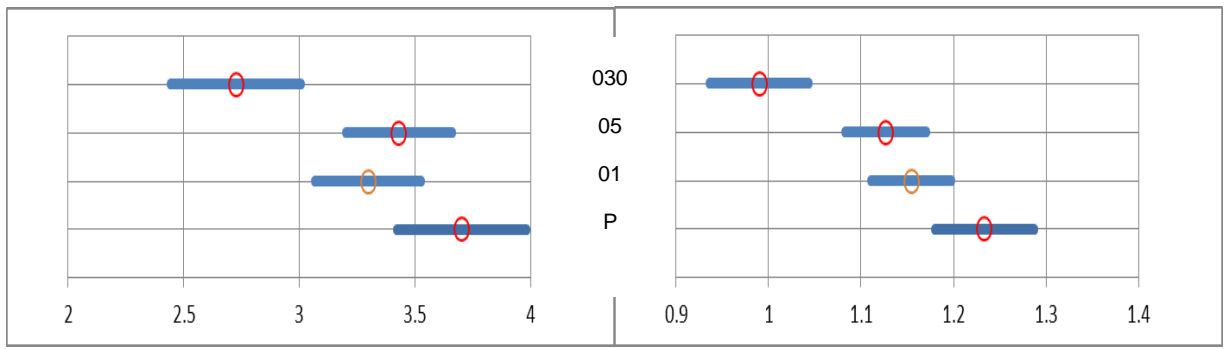

**Figure 6.** Estimated means of size of focal adhesions in cells on different substrates and confidence interval for the difference between substrates. A) long axis of focal adhesion and B) short axis of focal adhesion, alpha=0.01.

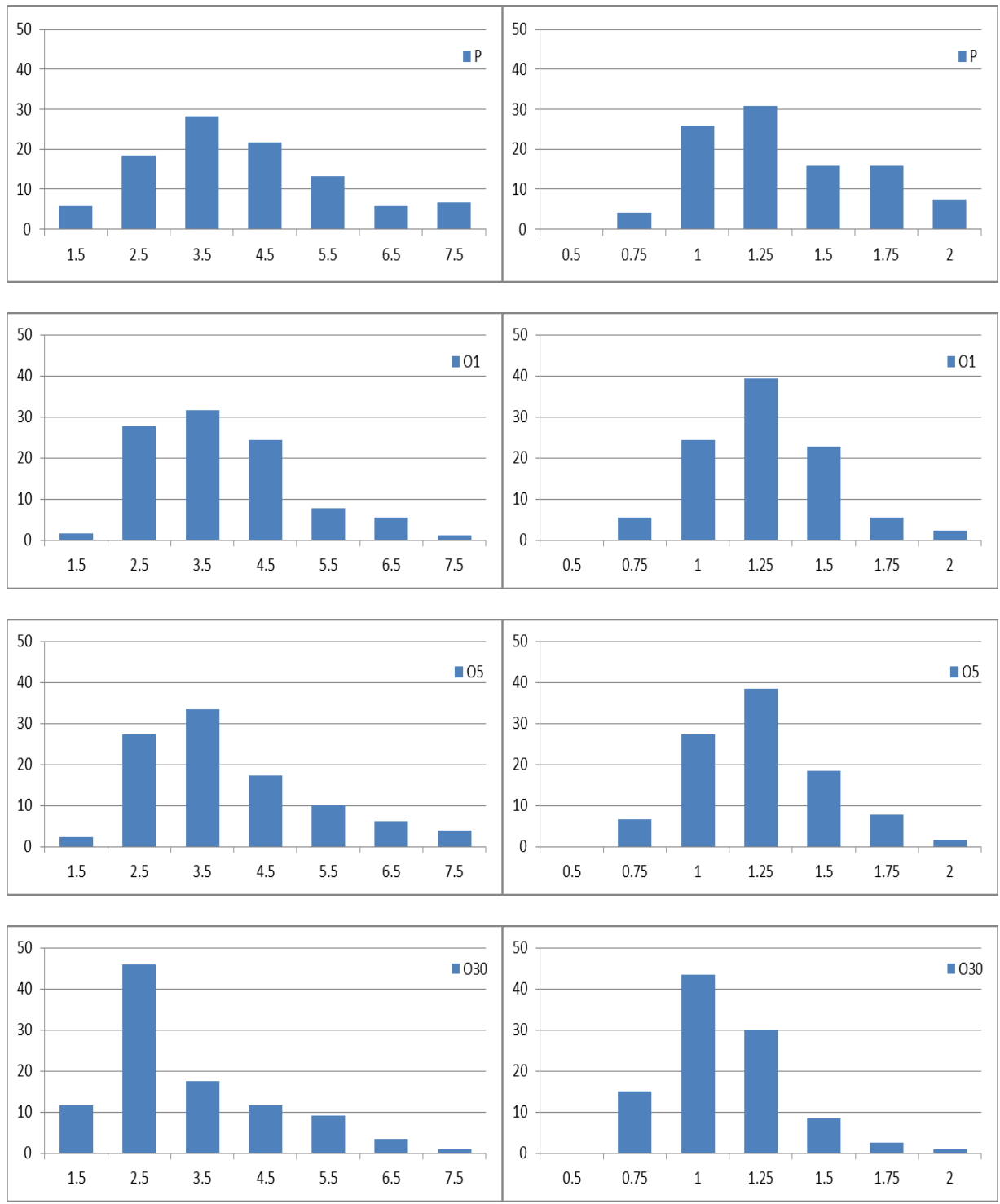

**Figure 7.** Relative frequency density histograms of size of focal adhesions in cells on different substrates. A) long axis of focal adhesion and B) short axis of focal adhesion.